\documentstyle[12pt]{article}
\input psfig
\long\def\comment#1{}
\begin{document}
\title{Quantum Information in a Distributed Apparatus}
\author{Subhash C. Kak\\
Department of Electrical \& Computer Engineering\\
Louisiana State University\\
Baton Rouge, LA 70803-5901; {\tt kak@ee.lsu.edu}}
%\date{20 December, 1996}
\maketitle

\begin{abstract}
We investigate the 
information provided about a specified distributed apparatus
of $n$ units
in the measurement of a quantum state. 
It is shown that, in contrast to such measurement of a classical
state, which is bounded by $\log (n + 1)$ bits, the information in a
quantum measurement is bounded by $3.7\times  n^{\frac{1}{2}}$ bits.
This means that the use of quantum apparatus 
offers an exponential advantage over classical apparatus.
\newline
{\bf Foundations of Physics, 28, 1998}
%\newline PACS numbers: 03.65.Bz, 42.50.Dv, 89.70.+c
%\newline Keywords: Quantum information, measurement, quantum Zeno effect
\end{abstract}

\section{Introduction}
Observers play a central role in quantum mechanics but there exists
no measure of intrinsic information associated with a quantum state.
The notions from classical information theory cannot be applied
directly here because of the fundamental uncertainty
in the quantum description.
For example, after the spin of a spin-$\frac{1}{2}$
particle has been measured to be in a particular direction, there
is still the probability $cos^{2} \theta/2$
that it will have a spin in a new direction at an angle of
$\theta$ from the previous one.
Or a photon that has been polarized in a certain direction will
still have a certain probability of being found in other directions.
Each such measurement, thus, provides new information.
This phenomenon of endless
information is the flip side of uncertainty.

Nevertheless,
the question of information can be made 
meaningful if
it is connected to the nature of the measurement apparatus.
Quantum interference experiments on single photons or other
particles reveal that the results give us information about
the measuring apparatus; the photon `senses' the entire 
measuring apparatus and it cannot be considered to be localized [3-8].
It is due to this property that interaction-free measurements can
be performed\cite{El93,Kw95}.
It stands to reason that if the measurement system is more
intricate than an interferometer, this increased complexity will
be correspondingly reflected in the measurement.

We consider the situation when the apparatus is
linearly distributed.
We ask the question whether the last measurement in this
apparatus can reveal to us information regarding the
previous stages of the apparatus.
In other words, we wish to speak not of any {\it intrinsic}
information of the object, but rather of the change
caused by the object in the information
of the classical distributed apparatus.

\section{Bits in a qubit}
We first look at the information that can be extracted from
a photon.
A qubit is a quantum state of the form
\begin{equation}
 |\phi\rangle = \alpha |0\rangle + \beta |1\rangle 
\end{equation}
where $\alpha, \beta$ are in general complex numbers and
$||\alpha||^2 + ||\beta||^2 = 1.$ The information that can
be extracted from a qubit depends on the measurement apparatus.
One may ask: What is the maximum information in bits that can be
obtained from a qubit?
We do not answer this question directly, but show how the
information in a qubit can depend on the measurement apparatus
and present a scheme where it can exceed two bits.

For convenience, we assume that the qubits are coded in the
polarization of photons.
The states $|0\rangle$ and $|1\rangle$ represent horizontally and vertically
polarized photons, respectively.
The information exchange protocol may be defined by the transmission,
according to a clock, of photons, which are detected
using appropriate polarizing filters. It is quite clear
that at each tick of the clock one could easily look for one of
at least three possibilities

\begin{equation}
|0\rangle, |1\rangle, ~no~photon. 
\end{equation}

This can be accomplished by a detector which is a horizontally
polarizing filter that is followed by a photo-detector $D_1$.
The polarizing filter is itself hooked to a measurement apparatus $D_2$
which can determine whether the filter has absorbed any photons.
This is shown in Figure 1.

%\fbox{Figure 1 about here}
\vspace{2mm}
\begin{figure}
\hspace*{0.5in}\centering{
\psfig{file=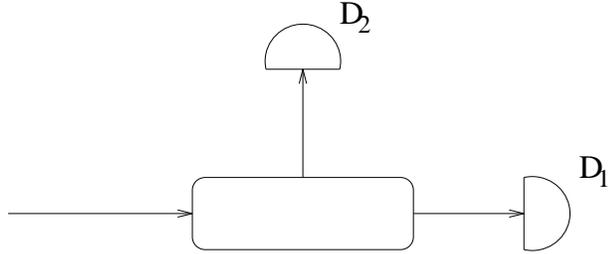,width=8cm}}
\caption{Measurements with two detectors $D_1$ and $D_2$}
\end{figure}

%\vspace{2mm}
If we represent the two $D_i$ measurements as a vector, we have
the following possibilites:

\begin{table}
\caption{Detector values and the corresponding states}
%\begin*{centering}
\begin{tabular}{||c|c||} \hline
$(D_1 , D_2)$ & state detection \\ \hline
(1, 0) & $|0\rangle$ \\ \hline
(0, 1) & $|1\rangle$ \\ \hline
(0, 0) & no photon \\ \hline
\end{tabular}
%\end{centering}
\end{table}

\vspace{3mm}
This means that the information being obtained at the receiver
is $ log_2 3~=~1.585$ bits per communication.
Since this implies equal probability of the three receiver states, this 
situation is the optimal with respect to the amount
of information.

Looking at the issue from the point of view of the $D_1 , D_2$
detectors, one might ask whether one can also detect the
remaining possibility, namely the entangled state
 \[|\nearrow\rangle =  \frac{1}{\sqrt 2} ( |0\rangle + |1\rangle ) \]

If one could do so, then the information in a qubit would
be equal to 2 bits. But this is not
possible 
since such an entangled state will
be detected either as $|0\rangle$ or $|1\rangle$.

\section{A distributed measurement apparatus}
The situation of a distributed apparatus is captured
most conveniently
in the realizations of the quantum Zeno effect\cite{Mi77,Pe80,Ag94}.
If a system is kept under constant observation, its state
does not change.
Consider a distributed apparatus consisting of $n$
horizontal polarizers (HP) 
interleaved with an equal number of 
polarization rotators (R), where each such
rotator shifts the polarization through
an angle of $\frac{\pi}{2n}$ 
(Figure 2). The measurement is made at the detector $D$
at the right of the interleaved
sequence of $R$s and $HP$s, where the intensity of the photons passing through
is measured. 

\begin{figure}
\hspace*{0.5in}\centering{
\psfig{file=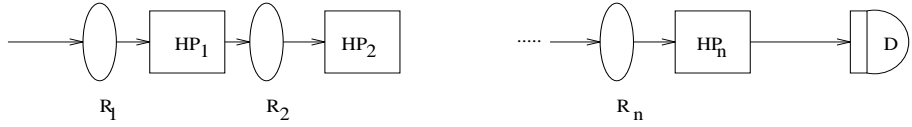,width=12cm}}
\caption{A distributed measurement apparatus: $n$ polarization rotators (R)
interleaved with horizontal polarizers (HP)}
\end{figure}

%\fbox{Figure 2 about here}

The probability that the photons will pass through each horizontal polarizer
is $cos^{2}\frac{\pi}{2n}$ of the previous stage.
Therefore, the output
at the end of the chain of $n$ detectors is:

\begin{equation}
 (cos^{2}\frac{\pi}{2n})^{n} \approx (1 - \frac{\pi^2 }{4 n^2}). 
\end{equation}

As $n \rightarrow \infty$, the output becomes 1.
This shows how a system that is continually observed does not change
its state.
But what happens if one or more of the horizontal polarizers (HP) is
missing?

Consider first that we deal with a fixed $n$ of HP.
The number of choices associated with the presence or absence
of HPs is $2^n$.
Each such choice may be represented by a binary vector, where $0$
represents the absence of an HP and $1$ represents its presence.
The information associated with each one of
these choices is $\log_2 (2^n) = n~ $bits.

\subsection{Classical measurements}
How much of information is conveyed by the measurements at D?
This will depend on the spectrum of light intensities received.
In the classical case we do not consider the rotators of Figure 1,
and we further take it
that each detector reduces the intensity by a
factor of $\alpha$; this is to maintain a parallel with the
quantum case where also the intensity is reduced.
If $k$ measurement units are on,
the intensity reaching the detector at
the right is $\alpha^k$.
The intensities at the detector will fall into a total of $n+1$ classes,
based on whether 0, 1, 2, ..., or $n$ units are on.
The $2^n$ states distribute into $n+1$ classes
according to the binomial distribution 
\begin{equation}
2^n = \sum_{k=0}^n
\left( \begin{array}{c}
    n \\  k
   \end{array} \right) .
\end{equation}

The information obtained at the detector can be
computed from the expression
$H(n) = \sum- p_i \log_2 p_i$ bits, where $p_i$ represents the 
probability of the class $i$.
%A tabulation for $n$ from 2 to 8 is given in Table 1.

%\begin{table}
%\caption{Information in a classical distributed apparatus for $n= 2,\cdots,8$}
%\begin{centering}
%\begin{tabular}{||l|r|r|r|r|r|r|r||} \hline
%n & 2& 3 & 4 & 5 &6 &7&8 \\ \hline
%Information & 1.5 & 1.8 & 2.02 & 2.2 & 2.33 & 2.43 & 2.53 \\ \hline
%\end{tabular}
%\end{centering}
%\end{table}

This information is bounded by $\log_2 (n+1)$ bits and as $n$ increases,
the difference between this bound and the actual value becomes
progressively worse because the light intensities do not
map uniformly with respect to the measurement units.

\subsection{Quantum measurements}

In the quantum case, each detector causes an intensity reduction
which is $cos^2 \theta$, where $\theta$ equals $\pi /2n$ times the
number of missing detectors in the sequence preceding it. 
The computation of intensity, therefore, is based on the count of
the groups of such missing units; if there is no missing unit, then
we count by $1$s. Clearly, this procedure implies a partitioning
of $n$ into its component parts. Furthermore, since the presence or
absence of a detector at the $nth$ location makes no difference
to the final result  (as in its absence the detector $D$ performs
the identical operation), each of the partitions will be counted twice.
The number of partitions of $n$ equals $p(n)$
(Table 2). 
The partition function $p(n)$ is well known in number theory.

Let the classes of partitions of $n$ be consecutively listed from
$1,\cdots,k$. 
Let
$a(k)$ be the number of elements in the class $k$. Since the total 
number of choices is $2^n$, we have

\begin{equation}
 \sum_{k=1}^{p(n)} a(k) = 2^n .
\end{equation}

\begin{table}
\caption{The partition function $p(n)$ with respect to $n$}
%\begin*{centering}
\begin{tabular}{||l|r|r|r|r|r|r|r|r|r|r||} \hline
$n$ &1& 2& 3 & 4 & 5 &6 &7&8 &9&10 \\ \hline
$p(n)$ & 1&2&3&5&7&11&15&22&30&42 \\ \hline
\end{tabular}
%\end{centering}
\end{table}

The partition function 
$p(n)$ builds up exponentially as we see in Table 2.
The value of $p(100)$ is 190,569,292.

In the case of our quantum mechanical measurement apparatus,
we can list the apparatus states as binary sequences.
These get distributed into intensity values based on the
partitions of the number $n$. 

{\it Example 1.} Let $n=3.$ The spectrum of energy values at D corresponding
to the various apparatus states (AS) 
$000, 001, 010, 011, 100, 101, 110, 111$ consists of
\[cos^2 \pi/2, cos^2 \pi/2, cos^2 \pi/3~ cos^2 \pi/6  , cos^2 \pi/3~ cos^2 \pi/6  , \]
\begin{equation}
 cos^2 \pi/6~ cos^2 \pi/3  , cos^2 \pi/6~ cos^2 \pi/3  , (cos^2 \pi/6)^3 , (cos^2 \pi/6 )^3 .
\end{equation}

These eight values form four distinct sets.
In other words, the intensities are obtained by partitioning $n = 3$
into its four partitions $p(3)$ of 3, 2+1, 1+2, 1+1+1+1, each counted
twice.
This is shown in Table 3.

\begin{table}
\caption{Intensities in the quantum distributed apparatus detector D, where $n=3$}
%\begin*{centering}
\begin{tabular}{||l|r|r|r|r|r|r|r|r||} \hline
AS & 000 & 001 & 010 & 011 & 100 & 101 & 110 & 111 \\ \hline
Intensity & 0 & 0 & 3/16 & 3/16 & 3/16 & 3/16 & 27/64 & 27/64 \\ \hline
\end{tabular}
%\end{centering}
\end{table}

This forms three classes with probabilities of $1/4, 1/2,  1/4$.
The average information provided by this is 
$\sum- p_i \log_2 p_i = 1.5$~bits.
This example illustrates how the task of computing the intensity
spectrum is equivalent to determining the partitions of $n$.

The properties of the partition function are well known.
%There exist expressions which give estimates of $p(n)$.
It is known that the partition function $p(n)$ satisfies
the following property\cite{Lo82}:

\begin{equation}
\lim_{n \rightarrow \infty} \frac{ ln~p (n)}{n^\frac{1}{2}} = \pi \sqrt{2/3}
\end{equation}

For large $n$, we can use the approximation
 
\begin{equation}
\log_2 p(n) \approx n^\frac{1}{2} \pi \sqrt{2/3} \log_2 e .
\end{equation}

Since the partitions define the number of classes of the
intensity spectrum at the detector,
we can compute the average information $H(n)$ by using the actual
probabilities for the classes.
This is shown in Figure 3.
If all of these classes were equally likely, then the information is

\begin{equation}
H(n) \leq 3.7 n^\frac{1}{2} ~ bits.
\end{equation}

Since the equal probability case represents the maximum entropy,
the relation (9) is an upper bound for the measurement arrangement
described in the paper.

For convenience, this relation may be written as

\begin{equation}
H(n) \approx n^\frac{1}{2}~ bits. 
\end{equation}

\begin{figure}
\psfig{file=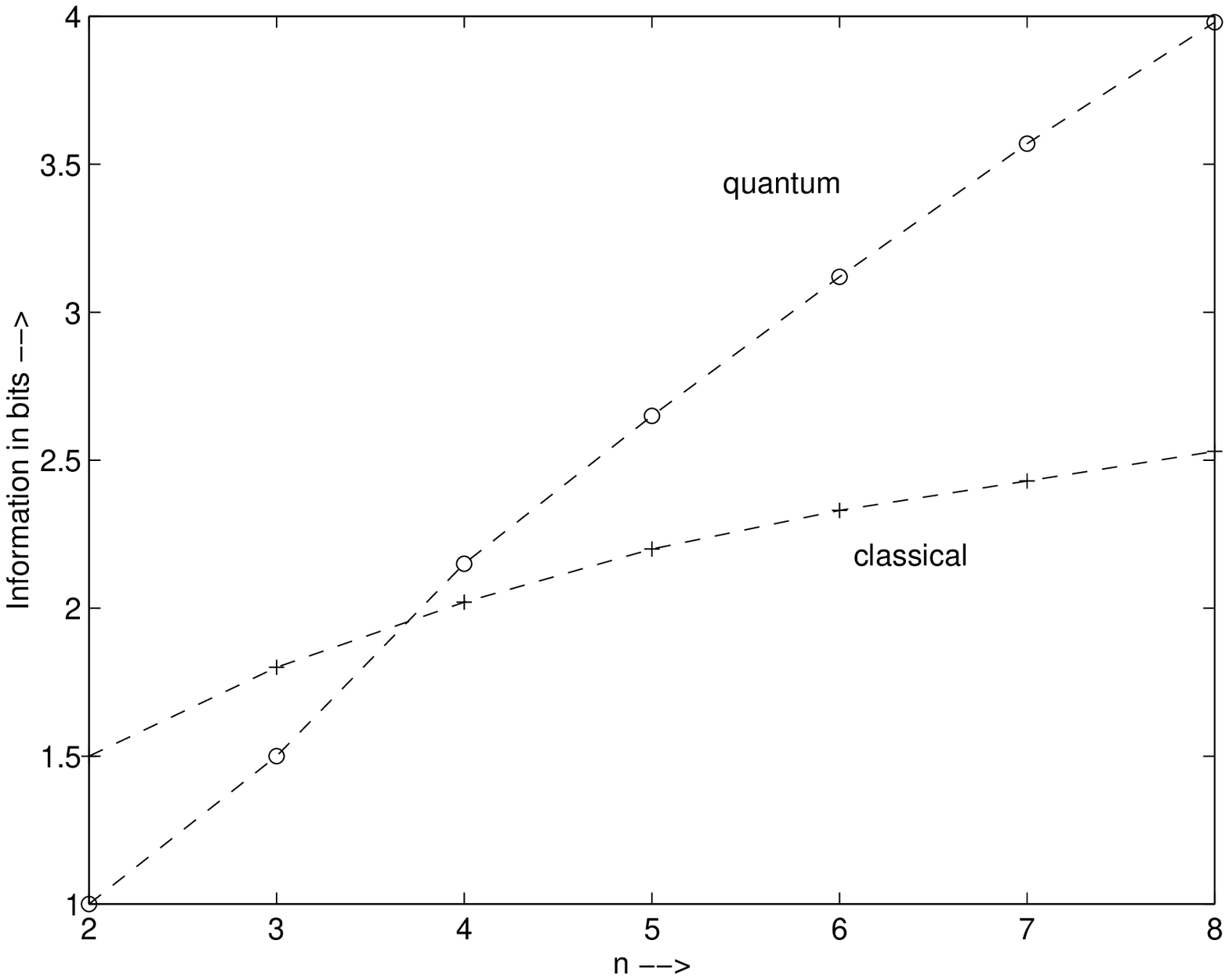,width=5in}
\vspace*{6in}
\caption{Classical {\em vs.\/} quantum information for small $n$\label{fig:inf}}
\end{figure}
%\fbox{Figure 3 about here}

Since, the class membership is based on the combinations of the
different partitions of $n$, rather than the combinations of $n$
directly as in the binomial distribution, the discrepancy between
the actual information value and its upper bound is much less
in the quantum case as compared to the classical case.
\section{Discussion}

The analysis done in this paper shows that a quantum
measurement gives information roughly
\begin{equation}
\sqrt{n} /\!\log n
\end{equation}
times that provided by the corresponding classical 
measurement, as shown in Figure 3.

As $n$ becomes large, the advantage over classical measurement
becomes enormous. 
Put differently, for the same performance, one would need
of the order of $2^{\sqrt{n}}$
classical elements where only $n$ quantum elements would suffice.
This exponential advantage is what makes quantum algorithms 
for solving problems faster than the corresponding
classical algorithms.
This is also the explanation for why, from a computational
point of view, quantum interactions are
so fundamentally different from classical interactions.

This paper has considered
the nature
of the measurement process 
from the point of view of intrinsic information.
Another perspective is to study situations which depart
from
classical logic;
these include the
Aharonov-Bohm, the EPR,
and the Hansbury Brown-Twiss effects\cite{Si95}. 
%according to which
%the diffraction of charged particles can be
%influenced by electromagnetic potentials under conditions where
%the electromagnetic fields are null.
A setting for 
detection of objects without any interactions with the object
was presented
by Elitzur and Vaidman\cite{El93}
with the
Mach-Zehnder particle interferometer as the measurement apparatus.
When both paths are clear, the particles are registered in
one detector only. But
if one of the paths is blocked then there is a fifty percent
chance that the other detector will receive the particle. If this happens,
then clearly one has determined the blockage of the path 
without having interacted with the object of the blockage.
Using the notion of distributed
measurement\cite{Kw95}, the percentage for detecting without
interaction can be further increased.
If the photons are transmitted over two paths rather than one,
we increase the variables that can be changed.

The research reported in this paper suggests several new
investigations.
One may wish to examine apparatus
which is distributed in more than one spatial dimension
and
study the
use of entangled states as well as that of
coupled EPR particles.

\end{document}